\documentclass{aa}
\usepackage{graphicx}
\usepackage{txfonts}

\usepackage{lscape}
\usepackage{placeins}
\usepackage{tabularx}
\usepackage[normalem]{ulem}

\defcitealias{sonnenfeldStatisticalStrongLensing2021}{Paper I}
\defcitealias{sonnenfeldStatisticalStrongLensing2022}{Paper III}
\defcitealias{sonnenfeldStatisticalStrongLensing2022a}{Paper IV}
\def\mfive{M_{\mathrm{DM},5}}

\usepackage[colorlinks=true,citecolor=blue,linkcolor=blue,urlcolor=blue]{hyperref}

\usepackage{orcidlink}

\begin{document}

\title{Statistical Strong Lensing\\ V. Inference with Microlensed Quasars}

\author{
Shuo Zhang\inst{1,2,3}\corrauth{zhangshuo01@sjtu.edu.cn}\orcidlink{0009-0002-4156-3718}
\and
Alessandro Sonnenfeld\inst{1,2,3}\corrauth{sonnenfeld@sjtu.edu.cn}\orcidlink{0000-0002-6061-5977}
}

\institute{
$^{1}$Department of Astronomy, School of Physics and Astronomy, Shanghai Jiao Tong University, Shanghai 200240, China\\
$^{2}$State Key Laboratory of Dark Matter Physics, School of Physics and Astronomy, Shanghai Jiao Tong University, Shanghai 200240, China\\
$^{3}$Key Laboratory for Particle Physics, Astrophysics and Cosmology, Ministry of Education, Shanghai Jiao Tong University, Shanghai 200240, China
}

\date{}

\abstract
{Microlensing of strongly lensed quasars provides direct information on the stellar component of lens galaxies, and can help break the degeneracy between the stellar initial mass function (IMF) and the dark matter. 
However, a statistical framework to forward model microlensing data while accounting for strong lensing selection effects is still missing.
We incorporated microlensing effects into the hierarchical model of Paper III, and investigated to what extent flux measurements of strongly lensed quasars can break the degeneracy between the stellar IMF and the dark matter distribution.
We tested the method on a mock sample of 219 doubly imaged quasars, lensed by circularly symmetric lenses.
When microlensing effects are modelled, the inference using image fluxes recovers the input foreground population parameters and provides tighter constraints than the inference without flux constraints. The improvement is clearest in the \(\mu_\alpha-\mu_5\) plane, where \(\mu_\alpha\) describes the mean stellar population synthesis mismatch parameter \(\log\alpha_{\rm SPS}\) and \(\mu_5\) describes the mean projected dark matter mass within \(5\,{\rm kpc}\).
The 68\% credible interval on $\mu_\alpha$ shrinks from 0.16 dex to 0.09 dex, with the degeneracy between $\mu_\alpha$ and $\mu_5$ significantly reduced.
By contrast, fitting the image fluxes without accounting for microlensing leads to large biases on the inferred parameters.
This work is a first step towards an accurate modelling of microlensing data.
Application to real lens samples will require dropping the circular symmetry assumption and adopting more realistic models for the microlensing magnification.
}


\maketitle
\nolinenumbers

\section{Introduction}
Stellar mass is a fundamental parameter for understanding galaxy formation and evolution. A common method for estimating the stellar masses of galaxies is stellar population synthesis (SPS) modelling \citep{conroyModelingPanchromaticSpectral2013}. In this approach, one compares observed spectral energy distributions (SEDs) with model predictions to infer the stellar mass of a galaxy, under assumptions about the initial mass function (IMF), star formation history, and stellar evolution. In particular, the inferred stellar masses depend sensitively on the assumed IMF, which remains a major uncertainty because its form is still poorly constrained. This motivates the need for independent constraints on the IMF. Strong gravitational lensing provides such an opportunity. Combining accurately measured image positions and flux ratios with SPS-based stellar masses can be used to calibrate the IMF of lens galaxies \citep{sonnenfeldStatisticalStrongLensing2021}. However, without flux information, it is difficult to simultaneously constrain the IMF and the inner dark matter density profile \citep{sonnenfeldSLACSStrongLens2025}.

Due to their compact emission regions, lensed quasars are often well approximated as point sources, allowing their image positions and fluxes to be measured with high precision.
However, the same compact emission regions make their fluxes susceptible to perturbations from compact objects along the line of sight.
At optical and X-ray wavelengths, these perturbations are dominated by stellar microlensing from stars in the foreground lens galaxy. 
This implies that the flux distribution in a statistical sample of strongly lensed quasars contains information on the stellar IMF of the galaxy population \citep{vernardosMicrolensingFluxRatio2019}.

In the coming years, wide-field surveys such as Euclid \citep{euclidcollaborationEuclidPreparationEuclid2022}, LSST \citep{ivezicLSSTScienceDrivers2019a}, and the Chinese Space station Survey Telescope (CSST) are expected to discover on the order of $10^5$ strong lenses, including a substantial population of lensed quasars \citep{oguriGravitationallyLensedQuasars2010,collettPOPULATIONGALAXYGALAXY2015, euclidcollaborationEuclidQuickData2026, csstcollaborationIntroductionChineseSpace2026}. These large datasets can provide the statistical power necessary to account for the perturbations introduced by microlensing, transforming what is typically a stochastic nuisance into a tractable component of the population inference.
However, as we explain below, we are still missing a robust method to model such data. 

There are two main approaches to modelling microlensed quasars. The first is to model the light curves of individual images, by simulating microlensing from different stellar configurations, and comparing the resulting mock light curves with the observations \citep{kochanekQuantitativeInterpretationQuasar2004}.
This method can in principle provide tight constraints, but it requires long-term and observationally expensive monitoring, as the characteristic timescale of microlensing variability is typically of order years \citep{vernardosMicrolensingStronglyLensed2024}.
The second approach consists in modelling microlensing statistically, using single-epoch observations from a large sample of systems \citep{schechterQualitativeAspectsQuasar2004}. In this work, we focus on this second method.

The probability distribution of the magnification due to microlensing depends on the local stellar surface mass density, but also on the lensing shear and on the total projected mass density (or, similarly, that of the dark matter component) \citep{vernardosMicrolensingStronglyLensed2024}.
Estimating the latter is particularly challenging.
\cite{oguriStellarDarkMatter2014}, in their analysis of 15 single-epoch microlensing measurements, simplified the problem by interpreting the microlensing effect as a constraint on the stellar mass fraction at the Einstein radius, thereby neglecting the radial offsets between individual images.
\citet{schechterCALIBRATIONSTELLARMASS2014} analysed a similar sample, and obtained the projected surface mass density profile by assuming that it scales perfectly with the half-light radius and the velocity dispersion (or a proxy for it), thereby neglecting scatter.
More recently, \citet{weisenbachStarsCosmicScales2025c} showed that
single-epoch microlensed image fluxes of lensed Type Ia supernovae
can be used to constrain the IMF mismatch parameter
$\alpha_{\rm SPS}$, which is defined as the ratio between the true stellar mass of a galaxy and the stellar mass inferred from SPS modelling assuming a reference IMF. However, they asserted that the convergence $\kappa$ and shear $\gamma$ at the image positions can be known exactly, while in
practice these quantities are subject to uncertainties associated with
the mass-sheet degeneracy and the choice of lens model.
A more fundamental problem is that these analyses characterize ensembles of strong lens galaxies without explicitly incorporating the selection function that maps the underlying galaxy population into the observed lens samples. Their inferred population-level constraints therefore cannot straightforwardly be interpreted as constraints on the underlying galaxy population. 

A suitable framework for inferring the properties of galaxies from a sample of lenses is the statistical formalism introduced in \citet[][hereafter Paper III]{sonnenfeldStatisticalStrongLensing2022}, which accounts for lensing selection. In this work, we extend that formalism by adding a treatment for microlensing.
We test the method on simplified simulated populations of foreground galaxies and background sources, from which a complete subset of lensed quasars is drawn.
Although our experiment is designed with quasars as background sources, our framework can be equally applied to any population of lensed point-like objects, such as supernovae.

The rest of the paper is organized as follows.
Section~\ref{sec:theory} introduces the microlensing formalism and the selected-sample likelihood.
Section~\ref{sec:simulation} describes the mock lens sample, and Section~\ref{sec:inference} describes the inference model.
Section~\ref{sec:results} presents the parameter recovery tests.
We discuss the main limitations in Section~\ref{sec:discussion} and summarize the conclusions in Section~\ref{sec:conclusion}.

Throughout this paper, we assume a flat \(\Lambda\)CDM cosmology with
\(\Omega_{\rm M}=0.3\) and
\(H_0=70\,{\rm km\,s^{-1}\,Mpc^{-1}}\).
We express all masses in solar masses, physical distances in kpc, angular quantities in arcseconds, and flux densities in \(\mu{\rm Jy}\), and \(\log\) denotes base-10 logarithms.
 \section{Microlensing Effect}
\label{sec:theory}

Our goal is to infer the population properties of lens galaxies from an observed sample of lensed quasars.
We denote these population parameters by \(\eta\).
They describe the parent population distribution of foreground galaxy properties, such as stellar mass, dark matter mass, and density slope.
For the $i$-th observed lens system, we denote the data by $d_i$. These data include non-flux observables, such as image positions and lens galaxy properties, which mainly constrain the macroscopic lens model, as well as the observed image fluxes, which are affected by microlensing.

In general, the local convergence $\kappa$, the shear $\gamma$, the smooth matter fraction $s$, the typical mass scale of the compact objects, and the source light distribution determine the microlensing magnification distribution \citep{wambsganssProbabilityDistributionsMagnification1992,vernardosMicrolensingStronglyLensed2024}. However, in the point-source limit, the physical length scales become degenerate. Consequently, the probability density function of the absolute magnification $\mu$ depends only on the local macro-parameters, which we denote as $\mathrm{P}(|\mu| \mid \kappa, \gamma, s)$.

The basic structure of the model is as follows. A set of lens and source parameters $\psi$ determines the image positions, the macromodel magnifications, and the local lensing quantities $(\kappa,\gamma,s)$ at each image. These local quantities determine the microlensing magnification distribution. The microlensing distribution and the noise model then map the intrinsic source flux into a distribution of observed image fluxes. The selection model determines the probability that the system enters the observed sample.

For later use, we express the microlensing magnification distribution in terms of the magnification relative to the absolute macromodel magnification. We define this relative microlensing magnification as
\begin{equation}
    x
    =
    \frac{|\mu|}{|\bar{\mu}|},
    \qquad
    \bar{\mu}
    =
    \frac{1}{(1-\kappa)^2-\gamma^2}.
\end{equation}
Here \(\bar{\mu}\) is the magnification predicted by the macromodel at the image position. The quantity \(\mu\) is the total microlensed magnification of a
macro-image, obtained by summing the flux magnifications of the unresolved micro-images produced by stellar microlenses.
For a fixed macromodel image, microlensing produces a distribution of possible total magnifications because different stellar microlens configurations lead to different flux perturbations.
By flux conservation, averaging over possible microlens configurations gives
\(\langle x\rangle = 1\), or equivalently
\(\langle |\mu| \rangle = |\bar{\mu}|\).

For image \(j\), the distribution of \(x_j\) is determined by the local
lensing quantities \(\kappa_j\), \(\gamma_j\), and \(s_j\), which are
themselves determined by the lens and source parameters \(\psi\).
We therefore write
\begin{equation}\label{eq:microlensing_single_image}
    \mathrm{P}(x_j\mid\psi)
    =
    \mathrm{P}\left(
    x_j
    \mid
    \kappa_j(\psi),
    \gamma_j(\psi),
    s_j(\psi)
    \right).
\end{equation}

The microlensing magnifications of different macro-images are independent at fixed \(\psi\). Therefore, the joint distribution of the relative microlensing magnifications
\(\vec{x}=(x_1,\ldots,x_{N_{\rm img}})\) is
\begin{equation}\label{eq:microlensing_joint}
    \mathrm{P}(\vec{x}\mid\psi)
    =
    \prod_{j=1}^{N_{\rm img}}
    \mathrm{P}(x_j\mid\psi).
\end{equation}

For a given realization of \(\vec{x}\), the image flux before measurement
noise is
\begin{equation}
    f_j
    =
    f_s|\bar{\mu}_j|x_j.
\end{equation}
The flux \(f_j\) is the summed flux of the unresolved micro-images
associated with macro-image \(j\).
Thus, \(\vec{x}\) acts as a set of latent variables that maps the
macroscopic lens model and the intrinsic source flux to the image fluxes.

The observed lens sample is a selected subset of the parent lens--source population.
We denote by $S$ the event that a lens-source pair satisfies the survey selection criteria and enters the observed sample. In general, the probability of satisfying $S$ depends on the lensing observables (e.g., image positions, image fluxes, or lens galaxy light) together with the corresponding observational noise model.

We consider an idealized survey in which all lensed quasars are found and included in the sample if all lensed images are brighter than the survey flux limit; that is, we assume no additional incompleteness from the lens finding procedure.
For a lens system with image fluxes \(\vec{f}\), the probability of it being selected given the observed flux \(\tilde{\vec{f}}\) is
\begin{equation}
    \mathrm{P}(S\mid\vec{f})
    =
    \int
    \mathrm{d}\tilde{\vec{f}}\,
    \mathrm{P}(S\mid\tilde{\vec{f}})
    \mathrm{P}(\tilde{\vec{f}}\mid\vec{f}).
\end{equation}
In the following, we assume that the source population and the selection function are known.

For a lens system in the observed sample, the likelihood is the probability of observing data \(d_i\) for a selected system, given \(\eta\):
\begin{equation}
    \label{eq:selected_likelihood}
    \mathrm{P}(d_i\mid\eta,S)
    =
    \frac{\mathrm{P}(d_i,S\mid\eta)}
    {\mathrm{P}(S\mid\eta)} .
\end{equation}
The term \(\mathrm{P}(d_i,S\mid\eta)\) gives the probability that a draw from the parent population enters the selected sample and has observed data \(d_i\). The term \(\mathrm{P}(S\mid\eta)\) is the population selection probability that selection occurs.

The quantity \(\mathrm{P}(d_i\mid\eta,S)\) is the probability density for a system in the selected catalogue to have observed data \(d_i\).
Using this likelihood, we still infer the parameters \(\eta\) of the underlying parent foreground galaxy population; the effect of selection on the distribution of the observed sample is included in the likelihood.
If one instead used \(\mathrm{P}(d_i\mid\eta)\), the parent population distribution without accounting for selection, to model the observed lens sample, the model would interpret the differences introduced by the selection function as intrinsic properties of the galaxy population, leading to a biased inference.

We separate the observed image fluxes from the remaining observables as
$d_i=(d_i^{(-\vec{f})},\vec{f}_i^{\,\rm obs})$.
With these definitions and selection model, we write the joint probability \(\mathrm{P}(d_i,S\mid\eta)\) as
\begin{equation}
    \label{eq:joint_selected_data}
\begin{aligned}
    \mathrm{P}(d_i,S\mid\eta)
    &=
    \int
    \mathrm{d}\vec{x}\,
    \mathrm{d}\psi\;
    \mathrm{P}(d_i^{(-\vec{f})}\mid\psi)
    \mathrm{P}(\vec{f}_i^{\,\rm obs}\mid\vec{f}(\vec{x},\psi))\\
    &\qquad\times\mathrm{P}(S\mid \vec{f}(\vec{x},\psi))
    \mathrm{P}(\vec{x}\mid \psi)
    \mathrm{P}(\psi\mid\eta) .
\end{aligned}
\end{equation}
The corresponding selection probability \(\mathrm{P}(S\mid\eta)\) is
\begin{equation}
    \label{eq:selection_probability}
\begin{split}
    \mathrm{P}(S\mid\eta)
    =
    \int
    &\mathrm{d}\vec{x}\,
    \mathrm{d}\psi\;
    \mathrm{P}(S\mid\vec{f}(\vec{x},\psi))
    \mathrm{P}(\vec{x}\mid\psi)
    \mathrm{P}(\psi\mid\eta).
\end{split}
\end{equation}
Table~\ref{tab:notation} lists the definitions of the quantities introduced in this section and later in the paper.

\begin{table*}[ht]
\caption{Notation used in this work.}
\label{tab:notation}

\begin{tabularx}{\textwidth}{@{\hspace{0.5cm}} c @{\hspace{2.5cm}} X}
\hline
\textbf{Symbol} & \textbf{Description} \\
\hline
$\boldsymbol{\psi}_g$ & Set of parameters describing a foreground galaxy \\
$\boldsymbol{\psi}_s$ & Set of parameters describing a background source \\
$\mathrm{P}_{\rm g}(\boldsymbol{\psi}_g)$ & Probability distribution of the foreground galaxy population \\
$\mathrm{P}_{\rm s}(\boldsymbol{\psi}_s)$ & Probability distribution of the background source population \\
$S$ & Event that a lens system satisfies the survey selection criteria and enters the observed sample \\
$\mathrm{P}(S\mid\eta)$ & Population selection probability \\
$\boldsymbol{\beta}$, $\beta$ & Source position in the source plane and its radial coordinate, \(\beta=|\boldsymbol{\beta}|\) \\
$\boldsymbol{d}_i$ & Array of observational data for the \(i\)-th lens system \\
$\boldsymbol{\eta}$ & Set of parameters describing the population of foreground galaxies \\
$M_*$ & True stellar mass of the lens galaxy \\
$M_{*,{\rm SPS}}$ & SPS-inferred stellar mass \\
$\alpha_{\rm SPS}$ & Stellar population synthesis mismatch parameter, \(M_*=\alpha_{\rm SPS}M_{*,{\rm SPS}}\) \\
$R_{\mathrm{e}}$ & Half-light radius of the stellar component \\
$\mfive$ & Projected dark matter mass enclosed within \(5\,{\rm kpc}\) \\
$\gamma_{\rm DM}$ & Slope of the dark matter density profile \\
$\theta$, $\theta_1,\theta_2$ & Angular coordinate in the image plane and coordinates of images 1 and 2 \\
$\bar{\mu}$ & Macromodel magnification at an image position \\
$x$ & Relative microlensing magnification \\
$f_s$ & Source intrinsic flux density \\
$\vec{f}$, $f_j$ & Image flux vector, where \(f_j\) is the total flux of the unresolved micro-images associated with image \(j\) \\
$\tilde{\vec{f}}$ & Auxiliary noisy image flux used only to model survey detection \\
$\vec{f}^{\,\rm obs}$ & Measured image fluxes entering the flux likelihood \\
$f_{\rm lim}$ & Limiting image flux density \\
$\kappa$ & Convergence at an image position \\
$\gamma$ & Shear at an image position \\
$s$ & Smooth matter fraction, \(s=1-\kappa_\star/\kappa\) \\
\hline
\end{tabularx}
\end{table*}
 \section{Simulation}
\label{sec:simulation}

In order to demonstrate how the formalism in the previous section works in practice, we generated a simulated sample of galaxy-scale strong lenses. As in the previous papers of this series \citep{sonnenfeldStatisticalStrongLensing2021,sonnenfeldStatisticalStrongLensing2022,sonnenfeldStatisticalStrongLensing2022a}, we assume that the foreground galaxies have axisymmetric mass distributions and that the background sources are point-like. The impact of these assumptions on our results is discussed in Section~\ref{sec:discussion}. 
In the following sections, we describe the foreground galaxy population, the background source population, the microlensing magnification model, and how we construct the selected lens sample.

\subsection{Foreground Galaxy Population}\label{ssec:fgd}

We describe the stellar component of each foreground galaxy with a de Vaucouleurs profile. We specify this component by the true stellar mass \(M_*\) and the half-light radius \(R_{\rm e}\). In addition to the true stellar mass, we introduce the SPS stellar mass \(M_{*,{\rm SPS}}\), which is the stellar mass inferred from SPS modelling under a fixed reference stellar IMF. We relate the true stellar mass \(M_*\) to \(M_{*,{\rm SPS}}\) through the SPS mismatch parameter \(\alpha_{\rm SPS}\):
\begin{equation}
     \alpha_{\rm SPS}
    =\frac
    { M_*}{M_{*,{\rm SPS}}}.
\end{equation}
Thus, \(\alpha_{\rm SPS}=1\) corresponds to the case in which the reference SPS model gives the true stellar mass.

We model the dark matter component as a spherical power-law density profile. We parameterize it by the projected dark matter mass enclosed within \(5\,{\rm kpc}\), denoted by \(\mfive\), and its logarithmic density slope by \(\gamma_{\rm DM}\).

For convenience, we define the logarithmic values of the dimensional physical quantities in the adopted units as
\begin{equation}
    \begin{split}
    &m_* \equiv \log \frac{M_*}{M_\odot}, \\
    &m_{*,{\rm SPS}} \equiv \log \frac{M_{*,{\rm SPS}}}{M_\odot}, \\
    &r_{\rm e} \equiv \log \frac{R_{\rm e}}{\mathrm{kpc}}, \\
    &m_5 \equiv \log \frac{\mfive}{M_\odot}.
    \end{split}
\end{equation}

We draw the SPS stellar mass from a Gaussian distribution in logarithmic space,
\begin{equation}
     m_{*,{\rm SPS}}
    \sim
    \mathcal{N}(10.9, 0.2^2).
\end{equation}
We choose this distribution as a simple approximation to the high-mass part of the quiescent galaxy stellar mass function at \(0.2<z<0.5\), motivated by the measurements of \citet{muzzinEVOLUTIONSTELLARMASS2013} under the assumption of a Chabrier IMF.

We then draw the SPS mismatch parameter from a Gaussian distribution in logarithmic space,
\begin{equation}
    \log \alpha_{\rm SPS}
    \sim
    \mathcal{N}(0.1, 0.2^2).
\end{equation}
This fixes the true stellar mass \(M_*\) for each foreground galaxy.

We draw the half-light radius from an SPS stellar mass--size relation:
\begin{equation}
    r_{\rm e}
    \sim
    \mathcal{N}\left(0.9 + 0.2(m_{*,{\rm SPS}} - 10.9), 0.1^2 \right).
\end{equation}

For the dark matter halo, we draw the projected mass within a \(5\,{\rm kpc}\) aperture from the following Gaussian distribution:
\begin{equation}
    m_5
    \sim
    \mathcal{N}\left(10.7 + (m_{*,{\rm SPS}} - 10.9), 0.15^2 \right).
\end{equation}

We draw the logarithmic slope of the dark matter density profile from a truncated Gaussian distribution:
\begin{equation}
    \gamma_{\rm DM}
    \sim
    \mathcal{N}\left(1.6 + 0.4(m_{*,{\rm SPS}} - 10.9), 0.1^2 \right),
    \quad
    1.2 < \gamma_{\rm DM} < 2.
\end{equation}

We choose this truncation so that, after combining the stellar component with the dark matter halo, the resulting mock galaxies remain close to the nearly isothermal total density regime considered in this work, while avoiding extremely shallow or steep dark matter inner slopes that may lead to numerical instabilities in the lensing calculations.

For simplicity, we place all foreground lenses at a fixed redshift $z_{\rm l}=0.3$.

\subsection{Background Source Population}

We describe each background source by its intrinsic flux density $f_s$, redshift $z_s$, and source plane position $\boldsymbol{\beta}$. For simplicity, we fix the source redshift to $z_s=2$.

We draw the intrinsic source flux density \(f_s\) from a log normal distribution,
\begin{equation}\label{sourcepop}
    \log\left(\frac{f_s}{\mu{\rm Jy}}\right)
    \sim
    \mathcal{N}(-1.92, 0.84^2).
\end{equation}
We adopt this broad phenomenological distribution as a computationally convenient
approximation to the source flux distribution implied by the bolometric quasar
luminosity function of \citet{shenBolometricQuasarLuminosity2020} at the adopted
source redshift, over the flux range most relevant for our selection calculation.

We assume that the distribution of the intrinsic source flux density \(f_s\) is known and follows the same form as that adopted in eq \eqref{sourcepop}.
For each foreground lens, we generate a random realisation of background
source positions drawn from a uniform distribution with an average number
density of \(n_{\rm bkg}=30\,{\rm arcmin}^{-2}\).
This value is substantially higher than the observed quasar surface density and is adopted solely to improve the efficiency of the mock generation.

\subsection{Microlensing Magnification Distribution Surrogate}
The microlensing magnification distribution cannot in general be written in closed analytical form. We therefore obtain it from numerical simulations, using the method of \citet{weisenbachEfficientGenerationMicrolensing2025} to generate microlensing magnification maps over the parameter range \((0<\kappa<8,\; 0<\gamma<8,\; 0<s<1)\).
We generate the maps assuming a mean compact object mass of \(1\,M_\odot\). Each map has a side length of \(50\,R_{{\rm E},\odot}\) and contains \(2000\times2000\) pixels.

\begin{figure}
    \centering
    \includegraphics[width=\linewidth]{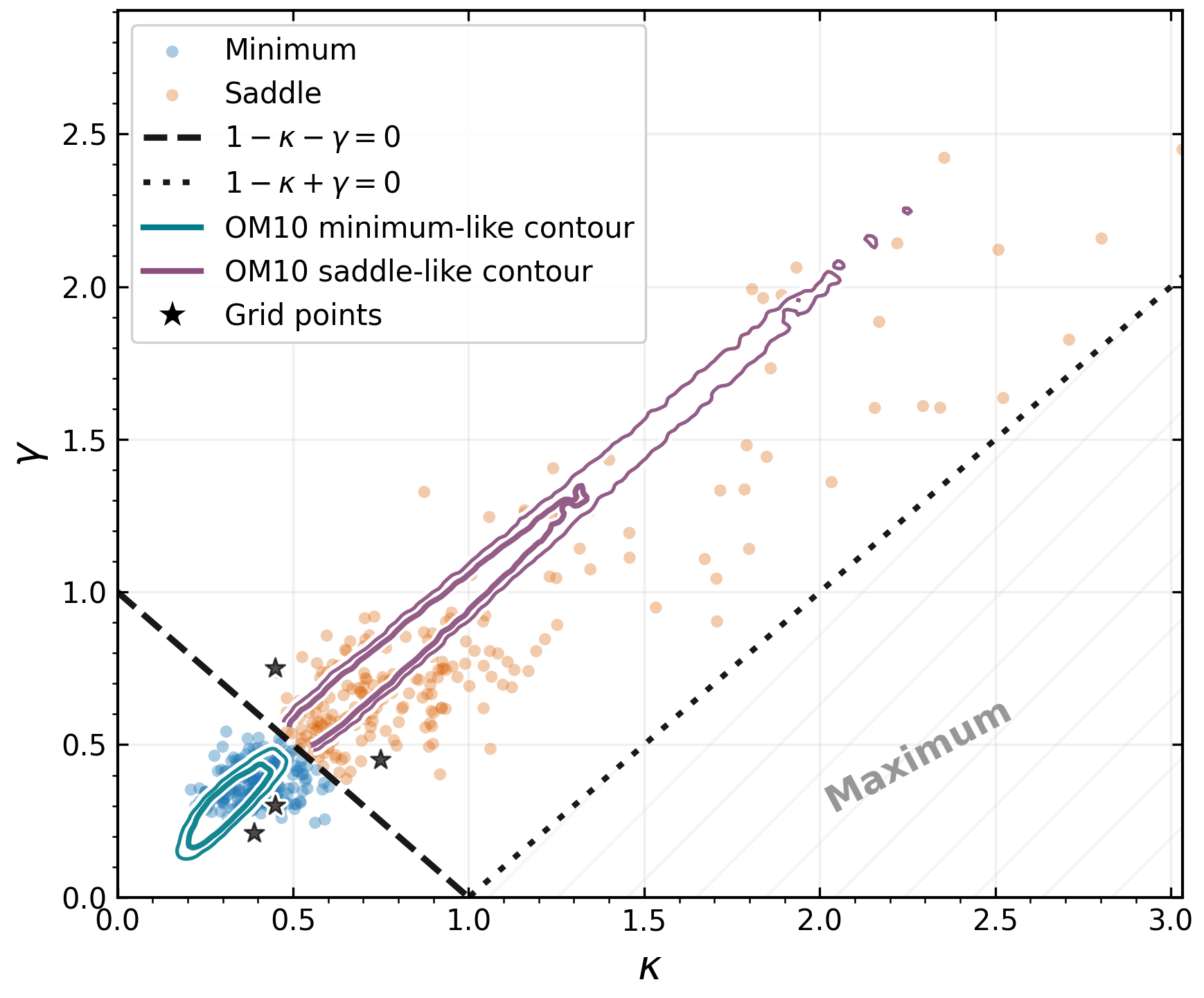}
    \caption{Local convergence \(\kappa\) and shear \(\gamma\) at the two image positions of the selected mock lenses. The blue and orange points show the minimum and saddle image branches. The dashed and dotted lines mark \(1-\kappa-\gamma=0\) and \(1-\kappa+\gamma=0\), respectively, which separate the main image parity regions relevant for the microlensing magnification distribution. The teal and purple contours show the corresponding regions occupied by OM10 lenses, and the black stars mark the grid points used for the example microlensing distributions in Figure~\ref{fig:mmd_surrogate}.}
    \label{fig:kgplane}
\end{figure}

Because we adopt the point source approximation, the microlensing magnification distribution depends only on the local lensing quantities \((\kappa,\gamma,s)\) and does not depend on the intrinsic source flux or on any source size parameter.
Figure~\ref{fig:kgplane} shows the local convergence and shear at the image positions of the selected mock lenses, together with the curves separating the main image parity regions.
The contours show the corresponding distributions for OM10 mock lenses \citep{oguriGravitationallyLensedQuasars2010} and indicate that our selected images occupy a broadly similar region of the \((\kappa,\gamma)\) plane.
The selected images lie within the \((\kappa,\gamma)\) range covered by the simulated microlensing maps.

\begin{figure}
    \centering
    \includegraphics[width=1\linewidth]{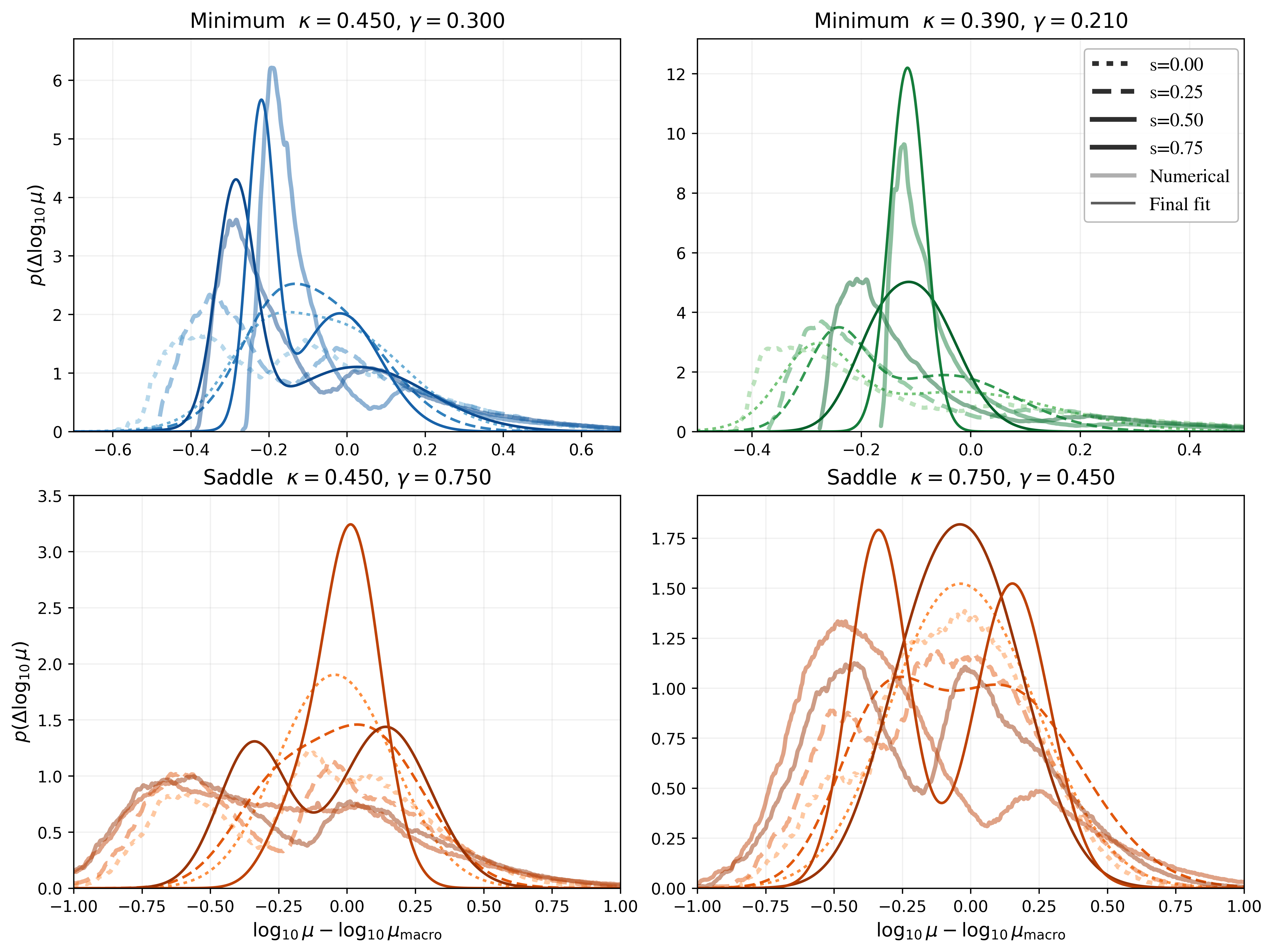}
    \caption{Examples of numerical microlensing magnification distributions and their surrogate fits. Each panel shows the distribution of \(\Delta\log\mu=\log|\mu|-\log|\bar{\mu}|=\log x\) for the indicated image parity and values of \((\kappa,\gamma)\). Different line styles correspond to different smooth matter fractions \(s\); pale curves show the numerical distributions and darker curves show the final surrogate fits.}
    \label{fig:mmd_surrogate}
\end{figure}

During the hierarchical inference, we must evaluate the microlensing magnification distribution repeatedly at different values of $(\kappa,\gamma,s)$. Directly using the numerical maps at every likelihood evaluation would be computationally expensive. We therefore use a surrogate model for the relative microlensing magnification distribution. Figure~\ref{fig:mmd_surrogate} shows examples of simulated distributions and surrogate fits. For fixed $(\kappa,\gamma,s)$, the surrogate approximates the numerical distribution of the relative magnification \(x\) in logarithmic space with a two-component Gaussian mixture,
\begin{equation}
\begin{aligned}
    \mathrm{P}(\log x\mid\kappa,\gamma,s)
    &=
    w\,\mathcal{N}(\log x\mid \mu_1,\sigma_1^2)
    \\
    &\quad+
    (1-w)\,\mathcal{N}(\log x\mid \mu_2,\sigma_2^2).
\end{aligned}
\end{equation}

Then we model the five mixture parameters
\begin{equation}
    \left\{
    w,\mu_1,\mu_2,\sigma_1,\sigma_2
    \right\}
\end{equation}
as smooth functions of the local lensing parameters \((\kappa,\gamma,s)\).
We construct the surrogate separately in different regions of the \((\kappa,\gamma)\) plane corresponding to different image parities. This allows the model to capture the qualitative changes in the shape of the microlensing magnification distribution between minimum and saddle images. In regions where the numerical distribution has only one main peak, the weight of the second Gaussian is driven toward zero, so that the model smoothly reduces to a single Gaussian. In regions where the numerical distribution is bimodal, both Gaussian components are active.

Although the surrogate does not reproduce the numerical microlensing magnification distributions exactly, it captures their principal characteristics, including the transition between unimodal and bimodal distributions, the peak locations, and the overall widths. We therefore use it as a computationally efficient approximation to the numerical microlensing magnification distributions in the mock test.

\subsection{Strong Lens Definition}

For each foreground galaxy, we solve the lens equation for background sources and identify the two main observable images of each multiply imaged source. We label them by \(j=1,2\), where image 1 is the outer positive-parity image and image 2 is the inner negative-parity image.
We ignore the possible central image, since it is highly demagnified for the centrally concentrated mass profiles considered in this work.

We include microlensing as a multiplicative perturbation to the macromodel magnification.
For each image, given the local convergence \(\kappa_j\), shear \(\gamma_j\), and smooth matter fraction \(s_j\), we generate a microlensing magnification and then compute the corresponding image fluxes \(f_1\) and \(f_2\).
We generate the observed fluxes by adding Gaussian measurement noise in logarithmic flux,
\begin{equation}
    \log f_j^{\rm obs}
    \sim
    \mathcal{N}(\log f_j,\epsilon_f^2),
    \qquad j=1,2 .
\end{equation}

In the mock sample, the selection event \(S\)
occurs when both observed image fluxes are above the limiting flux,
\begin{equation}
    S
    =
    \left\{
    f_1^{\rm obs}>f_{\rm lim},
    \quad
    f_2^{\rm obs}>f_{\rm lim}
    \right\}.
\end{equation}
We set \(f_{\rm lim}=0.145\,\mu{\rm Jy}\), corresponding to \(m_{\rm AB}=26\). For each foreground galaxy, we examine the generated background
sources sequentially until one produces a multiply imaged system
satisfying the flux selection criterion.

\begin{figure*}
    \centering
    \includegraphics[width=\textwidth]{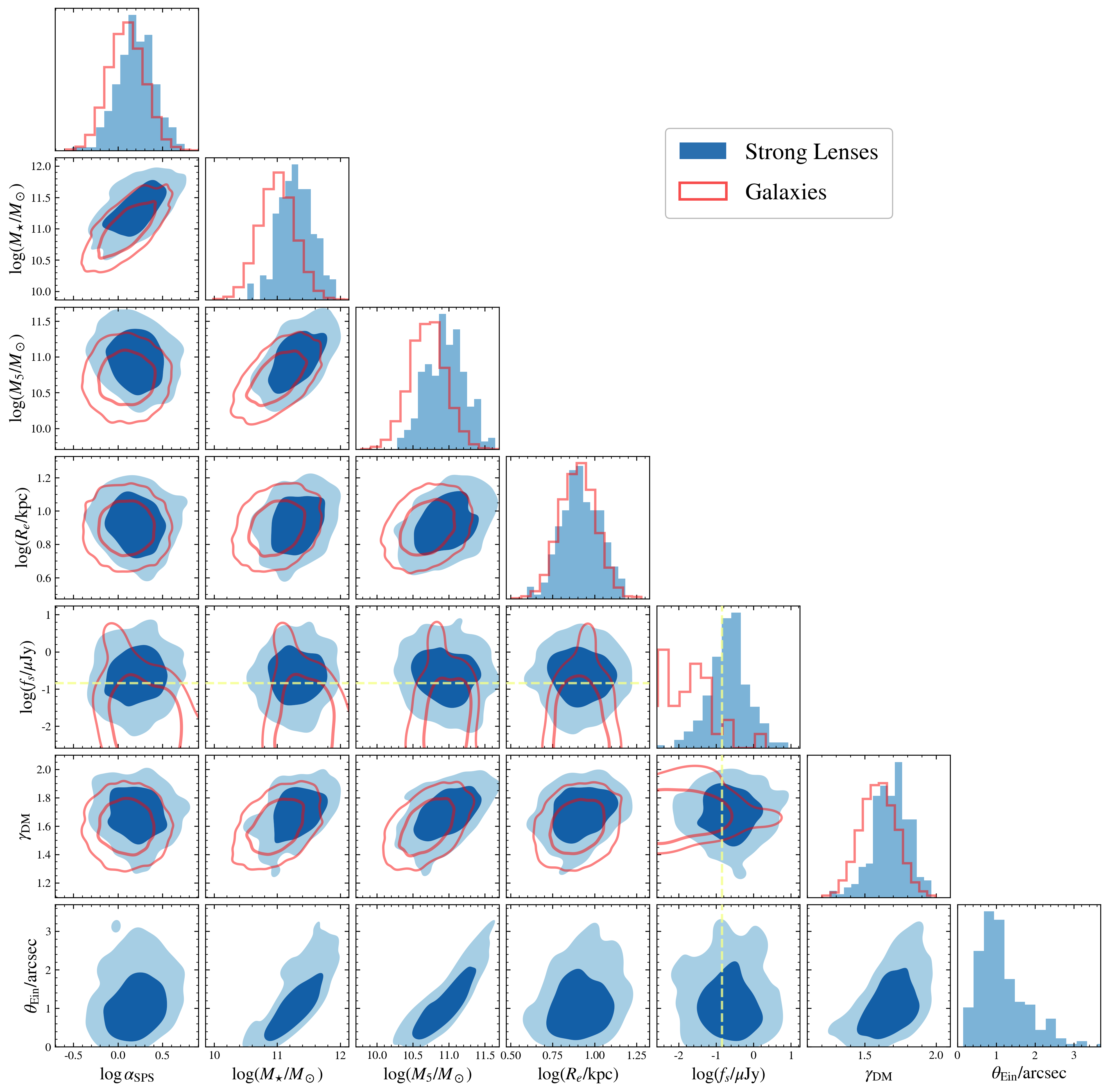}
    \caption{Distributions of the foreground galaxy, source, and lensing properties in the mock strong lens sample. Yellow dashed lines mark the flux density corresponding to \(m_{\rm AB}=26\).}
    \label{fig:mock_population}
\end{figure*}

Applying this procedure to the parent foreground galaxy sample produces 219 mock strong lenses out of 280000 foreground galaxies. Figure~\ref{fig:mock_population} shows the distribution in foreground galaxy properties, background source properties, and Einstein radius of the strong lenses. Compared to the general population, strong lens galaxies tend to have higher stellar and dark matter masses, as well as a slightly steeper inner dark matter slope. Table~\ref{tab:lens_params} summarizes the population and selection parameters used to generate the mock sample.

\begin{table*}[htbp]
\centering
\caption{Parameters used to generate the mock lens and source populations.}
\label{tab:lens_params}
\renewcommand{\arraystretch}{1.2}
\begin{tabular}{p{0.1\textwidth} p{0.6\textwidth} c}
\hline
\textbf{Symbol} & \textbf{Description} & \textbf{Value} \\
\hline
$\mu_\star$ & Mean of $m_{*,{\rm SPS}}$ & $10.9$ \\
$\sigma_\star$ & Scatter of $m_{*,{\rm SPS}}$ & $0.2$ \\
$\mu_\alpha$ & Mean of $\log \alpha_{\rm SPS}$ & $0.1$ \\
$\sigma_\alpha$ & Scatter of $\log \alpha_{\rm SPS}$ & $0.2$ \\
$\mu_{R,0}$ & Size--mass relation intercept & $0.9$ \\
$\beta_R$ & Size--mass relation slope & $0.2$ \\
$\sigma_R$ & Scatter of $r_{\rm e}$ & $0.1$ \\
$\mu_5$ & Dark matter mass relation intercept for $m_5$ & $10.7$ \\
$\beta_5$ & Dark matter mass relation slope & $1.0$ \\
$\sigma_5$ & Scatter of $m_5$ & $0.15$ \\
$\mu_\gamma$ & Inner dark matter slope relation intercept & $1.6$ \\
$\beta_\gamma$ & Inner dark matter slope relation slope & $0.4$ \\
$\sigma_\gamma$ & Scatter of $\gamma_{\rm DM}$ & $0.1$ \\
$z_{\rm l}$ & Lens redshift & $0.3$ \\
$z_{\rm s}$ & Source redshift & $2.0$ \\
$\mu_{\log f_s}$ & Mean of $\log(f_s/\mu{\rm Jy})$ & $-1.92$ \\
$\sigma_{\log f_s}$ & Scatter of $\log f_s$ & $0.84$ \\
$n_{\rm bkg}$ & Background source surface number density & $30\,{\rm arcmin}^{-2}$ \\
$f_{\rm lim}$ & Detection threshold on lensed image flux density, corresponding to $m_{\rm AB}=26$ & $0.145\,\mu{\rm Jy}$ \\
$\epsilon_f$ & Photometric uncertainty in $\log f$ & $0.04$ dex \\
$\epsilon_\star$ & Observational uncertainty of $m_{*,{\rm SPS}}$ & $0.1$ dex \\
\hline
\end{tabular}
\end{table*}
 \section{Inference}
\label{sec:inference}

We fit the simulated strong lensing data with a model that describes the
joint distribution of the galaxy population, the background source population,
and the survey selection model, generalizing the method introduced in
\citetalias{sonnenfeldStatisticalStrongLensing2022}. We adopt the same
galaxy and background source population models as those used to generate
the mock sample, while allowing the galaxy population parameters to vary.
We describe each aspect of the model in the following sections.

\subsection{Foreground galaxy population model}

We model each foreground galaxy with a stellar component that follows a de Vaucouleurs profile with half-light radius $R_{\rm e}$ and SPS stellar mass $M_{*,{\rm SPS}}$, and with a dark matter component that follows a spherical power-law density profile. We assume exact lens redshift measurements. As in Paper III, we denote with $\psi_{\rm g}$ the foreground galaxy parameters and $\psi_{\rm s}$ the background source parameters, with
$\psi=(\psi_{\rm g},\psi_{\rm s})$.

We parametrize each foreground galaxy by the SPS stellar mass
$M_{*,{\rm SPS}}$, the SPS mismatch parameter \(\alpha_{\rm SPS}\),
the effective radius \(R_{\rm e}\), the projected dark matter mass enclosed
within \(5\,{\rm kpc}\), \(\mfive\), and the dark matter density slope
\(\gamma_{\rm DM}\). The galaxy parameter set is therefore
\begin{equation}
    \psi_{\rm g}
    \equiv
    \left\{
    m_{*,{\rm SPS}},
    \log\alpha_{\rm SPS},
    r_{\rm e},
    m_5,
    \gamma_{\rm DM}
    \right\}.
\end{equation}

We assume the following probability distribution for the foreground galaxy parameters:
\begin{equation}\label{models}
\begin{split}
    \mathrm{P}_{\rm g}(\psi_{\rm g}\mid\eta)
    &=
    \mathcal{M}(m_{*,{\rm SPS}})
    \mathcal{A}(\log\alpha_{\rm SPS})
    \mathcal{R}(r_{\rm e})
    \\
    &\quad\times
    \mathcal{H}(m_5)
    \mathcal{Y}(\gamma_{\rm DM}) .
\end{split}
\end{equation}
Here \(\mathcal{M}\) and \(\mathcal{R}\) describe the distributions of the SPS stellar mass and effective radius of the foreground galaxies, respectively.
In principle, one should constrain these distributions from the observed \(M_{*,{\rm SPS}}\) and \(R_{\rm e}\) distributions of the parent foreground galaxy sample.
For simplicity, we assume that they are known exactly and fix them to the Gaussian forms specified above.
This is a reasonable approximation, since large galaxy samples can measure the stellar mass and size distributions with high precision \citep[see e.g.][]{sonnenfeldHyperSuprimeCamView2019}.

The term \(\mathcal{A}\) describes the distribution of the SPS mismatch parameter.
We model it as a Gaussian distribution in \(\log\alpha_{\rm SPS}\),
\begin{equation}
    \mathcal{A}(\log\alpha_{\rm SPS})
    \sim
    \mathcal{N}(\mu_\alpha,\sigma_\alpha^2).
\end{equation}

The term \(\mathcal{H}\) describes the distribution of the projected dark matter mass enclosed within \(5\,{\rm kpc}\).
We model it as a Gaussian distribution in $m_5$,
\begin{equation}
    \mathcal{H}(m_5)
    \sim
    \mathcal{N}
    \left[
    \mu_5+\beta_5
    \left(m_{*,{\rm SPS}}-m_{*,{\rm SPS}}^{\rm piv}\right),
    \sigma_5^2
    \right].
\end{equation}
Here $m_{*,{\rm SPS}}^{\rm piv}$ is the pivot value of the SPS stellar mass variable, which we set equal to the mean value of that variable in the parent foreground galaxy population, $m_{*,{\rm SPS}}^{\rm piv}=10.9$.
With this choice, \(\mu_5\) represents the mean value of $m_5$ for a galaxy with the typical SPS stellar mass of the sample, while \(\beta_5\) describes how the projected dark matter mass varies with stellar mass.

The term \(\mathcal{Y}\) describes the distribution of the dark matter density slope.
We model it as a truncated Gaussian distribution,
\begin{equation}
\begin{split}
    \mathcal{Y}(\gamma_{\rm DM})
    &\sim
    \mathcal{N}
    \left[
    \mu_\gamma+\beta_\gamma
    \left(m_{*,{\rm SPS}}-m_{*,{\rm SPS}}^{\rm piv}\right),
    \sigma_\gamma^2
    \right],
    \\
    &\qquad
    1.2<\gamma_{\rm DM}<2 .
\end{split}
\end{equation}
With the same pivot choice, \(\mu_\gamma\) represents the mean dark matter density slope at the typical SPS stellar mass of the sample, while \(\beta_\gamma\) describes the stellar mass dependence of the slope.

We control the foreground population model with the hyperparameter vector
\begin{equation}
    \eta
    =
    \left\{
    \mu_\alpha,\sigma_\alpha,
    \mu_5,\beta_5,\sigma_5,
    \mu_\gamma,\beta_\gamma,\sigma_\gamma
    \right\}.
\end{equation}
The goal of the inference is to recover \(\eta\) from the data.
The distribution \(\mathrm{P}_{\rm g}(\psi_{\rm g}\mid\eta)\) reduces to the true distribution used to generate the mock sample when
\begin{equation}
    \eta_{\rm true}
    =
    \left\{
    0.1,\,0.2,\,
    10.7,\,1.0,\,0.15,\,
    1.6,\,0.4,\,0.1
    \right\}.
\end{equation}

\subsection{Background Source Population Model}

We assume that redshifts of background sources are exactly measured.
We therefore describe each source by its intrinsic flux density \(f_s\) and radial source plane position \(\beta\):
\begin{equation}
    \psi_{\rm s}
    \equiv
    \left\{
    f_s,\beta
    \right\}.
\end{equation}

We assume that the distribution in \(\psi_{\rm s}\) of the background quasar population is known exactly from prior information.
We discuss possible strategies for relaxing this assumption in Section~\ref{sec:discussion}.
We therefore fix the probability distribution \(\mathrm{P}_{\rm s}\) to the one used to generate the mock sample in Section~\ref{sec:simulation}.
Since we fix the source redshift, we write the source population distribution as
\begin{equation}
    \mathrm{P}_{\rm s}(\psi_{\rm s})
    =
    \mathrm{P}(f_s)\,
    \mathrm{P}(\beta).
\end{equation}

We assume that the angular position of a background source on the sky is
uncorrelated with its other properties and is uniformly distributed over the
source plane. The corresponding distribution of the radial source position
\(\beta=|\boldsymbol{\beta}|\) is then
\begin{equation}
    \mathrm{P}(\beta)\propto\beta.
\end{equation}

\subsection{Selection Model}

The selection event \(S\) describes whether the observed image fluxes, which are measured with a log-normal uncertainty $\epsilon_f$, are sufficiently bright to enter the lens sample.
The corresponding selection probability is therefore
\begin{equation}
    \mathrm{P}(S\mid \vec{f})
    =
    \prod_{j=1}^{2}
    \frac{1}{2}
    \operatorname{erfc}
    \left[
    \frac{\log f_{\rm lim}-\log f_j}{\sqrt{2}\epsilon_f}
    \right].
\end{equation}

\subsection{Likelihood and Posterior Sampling}

We fit the population model described above to the mock strong-lens sample.
For each selected lens system, we use the following data:
(1) the positions of the two images of the lensed source,
\(\theta_{1,i}^{\rm obs}\) and \(\theta_{2,i}^{\rm obs}\), measured exactly;
(2) the effective radius of the foreground galaxy, \(r_{{\rm e},i}^{\rm obs}\), measured exactly;
(3) the SPS stellar mass of the foreground galaxy, \(m_{*,{\rm SPS},i}^{\rm obs}\), measured with uncertainty \(\epsilon_\star\);
and (4) the fluxes of the two lensed images, \(f_{1,i}^{\rm obs}\) and \(f_{2,i}^{\rm obs}\), measured with uncertainty \(\epsilon_f\).
The data vector for lens \(i\) is
\begin{equation}
    d_i
    =
    \left\{
    \theta_{1,i}^{\rm obs},
    \theta_{2,i}^{\rm obs},
    \log f_{1,i}^{\rm obs},
    \log f_{2,i}^{\rm obs},
    m_{*,{\rm SPS},i}^{\rm obs},
    r_{{\rm e},i}^{\rm obs}
    \right\}.
\end{equation}
In practice, we apply the uncertainties \(\epsilon_f\) and \(\epsilon_\star\) to the logarithmic fluxes and to the logarithmic SPS stellar mass, respectively.

For a sample of selected lenses, we write the posterior probability of the hyperparameters as
\begin{equation}
    \mathrm{P}(\eta\mid\{d_i\})
    \propto
    \mathrm{P}(\eta)
    \prod_i
    \mathrm{P}(d_i\mid\eta,S).
\end{equation}
For each selected lens, we use the likelihood defined in
Eq.~\eqref{eq:selected_likelihood}.
Specializing the joint probability in Eq.~\eqref{eq:joint_selected_data}
to the mock model gives
\begin{equation}
\begin{aligned}
    \mathrm{P}(d_i,S\mid\eta)
    =&
    \int
    \mathrm{d}\vec{x}\,
    \mathrm{d}\psi_{\rm g}\,
    \mathrm{d}\psi_{\rm s}\;
    \mathrm{P}\left(
        d_i^{(-\vec{f})}
        \mid
        \psi_{\rm g},\psi_{\rm s}
    \right)
    \\
    &\times
    \mathrm{P}\left(
        \vec{f}_i^{\,\rm obs}
        \mid
        \vec{f}(\vec{x},\psi_{\rm g},\psi_{\rm s})
    \right)
    \mathrm{P}\left(
        S
        \mid
        \vec{f}(\vec{x},\psi_{\rm g},\psi_{\rm s})
    \right)
    \\
    &\times
    \mathrm{P}\left(
        \vec{x}
        \mid
        \psi_{\rm g},\psi_{\rm s}
    \right)
    \mathrm{P}_{\rm g}(\psi_{\rm g}\mid\eta)
    \mathrm{P}_{\rm s}(\psi_{\rm s}) .
\end{aligned}
\end{equation}
Here \(\mathrm{P}(\vec{f}_i^{\,\rm obs}\mid\vec{f})\) denotes the measurement likelihood of the observed image fluxes, modeled as independent Gaussian errors in \(\log f\) with dispersion \(\epsilon_f\). The effects of stellar microlensing are incorporated into
\(\mathrm{P}(\vec{x}\mid\psi)\), as described in Eqs.~\eqref{eq:microlensing_single_image}--\eqref{eq:microlensing_joint}.

The non-flux part of the likelihood contains the exact constraints on the image positions and half-light radius:
\begin{equation}
\begin{aligned}
    \mathrm{P}(d_i^{(-\vec{f})}\mid\psi)
    =
    &
    \delta\left(r_{\rm e}-r_{{\rm e},i}^{\rm obs}\right)
    \mathrm{P}(m_{*,{\rm SPS},i}^{\rm obs}\mid m_{*,{\rm SPS}})
    \\
    &\times
    \delta\left(\theta_1(\psi)-\theta_{1,i}^{\rm obs}\right)
    \delta\left(\theta_2(\psi)-\theta_{2,i}^{\rm obs}\right).
\end{aligned}
\raisetag{1.2\baselineskip}
\end{equation}
The term
\(\mathrm{P}(m_{*,{\rm SPS},i}^{\rm obs}\mid m_{*,{\rm SPS}})\) denotes the Gaussian measurement likelihood for the observed SPS stellar mass, with dispersion \(\epsilon_\star\). Since we treat the image positions as exact, we use them to eliminate two latent variables.
For each value of the remaining lens parameters,
$m_{*,{\rm SPS}}$, $r_{\rm e}$, $m_5$, and \(\gamma_{\rm DM}\),
we solve the lens equation for the true stellar mass and source position that reproduce the two observed image positions.
We denote the resulting solution for lens \(i\) by
$(\hat{m}_{*,i},\hat{\beta}_i)$.
We then obtain the corresponding SPS mismatch parameter from
\(\log\hat{\alpha}_{{\rm SPS},i}
=
\hat{m}_{*,i}
-
m_{*,{\rm SPS}}\).

This change of variables introduces the Jacobian factor
\begin{equation}
    \left|
    \det J_i
    \right|
    =
    \left|
    \det
    \frac{
    \partial(m_*,\beta)
    }{
    \partial(\theta_1,\theta_2)
    }
    \right|_
    {(m_*,\beta)=(\hat{m}_{*,i},\hat{\beta}_i)} .
\end{equation}

After applying these exact constraints, we evaluate
\(\mathrm{P}(d_i,S\mid\eta)\) as an integral over the remaining variables:
\begin{equation}
\begin{aligned}
    \mathrm{P}(d_i,S\mid\eta)
    =
    \int
    &\mathrm{d}m_{*,{\rm SPS}}\,
    \mathrm{d}m_5\,
    \mathrm{d}\gamma_{\rm DM}\,
    \mathrm{d}f_s\,
    \mathrm{d}\vec{x}
    \\
    &\times
    \mathrm{P}\!\left(
    \vec{f}_i^{\,\rm obs}
    \mid
    \vec{f}(\vec{x},
    \hat{m}_{*,i},
    r_{{\rm e},i}^{\rm obs},
    m_5,
    \gamma_{\rm DM},
    f_s,
    \hat{\beta}_i)
    \right)
    \\
    &\times
    \mathrm{P}\!\left(
    S
    \mid
    \vec{f}(\vec{x},
    \hat{m}_{*,i},
    r_{{\rm e},i}^{\rm obs},
    m_5,
    \gamma_{\rm DM},
    f_s,
    \hat{\beta}_i)
    \right)
    \\
    &\times
    \mathrm{P}\!\left(
    \vec{x}
    \mid
    \hat{m}_{*,i},
    r_{{\rm e},i}^{\rm obs},
    m_5,
    \gamma_{\rm DM},
    f_s,
    \hat{\beta}_i
    \right)
    \\
    &\times
    \mathrm{P}_{\rm g}\!\left(
    m_{*,{\rm SPS}},
    \log \hat{\alpha}_{{\rm SPS},i},
    r_{{\rm e},i}^{\rm obs},
    m_5,
    \gamma_{\rm DM}
    \mid\eta
    \right)
    \\
    &\times
    \mathrm{P}_{\rm s}\!\left(
    f_s,\hat{\beta}_i
    \right)
    \left|\det J_i\right| .
\end{aligned}
\end{equation}

The corresponding selection probability for the population in Eq.~\eqref{eq:selection_probability} is obtained by averaging the selection probability over all possible foreground galaxies and background sources drawn from the population model:
\begin{equation}
\begin{split}
    \mathrm{P}(S\mid\eta)
    =
    \int
    &\mathrm{d}\vec{x}\,
    \mathrm{d}\psi_{\rm g}\,
    \mathrm{d}\psi_{\rm s}\;
    \mathrm{P}\!\left(
        S
        \mid
        \vec{f}(\vec{x},\psi_{\rm g},\psi_{\rm s})
    \right)
    \\
    &\times
    \mathrm{P}\!\left(
        \vec{x}
        \mid
        \psi_{\rm g},\psi_{\rm s}
    \right)
    \mathrm{P}_{\rm g}(\psi_{\rm g}\mid\eta)
    \mathrm{P}_{\rm s}(\psi_{\rm s}) .
\end{split}
\end{equation}

We sampled the posterior distribution of \(\eta\) using Markov chain Monte Carlo.
We adopted uniform priors for all components of \(\eta\) over broad ranges that contain the input values.
At each draw of \(\eta\), we computed both \(\mathrm{P}(S\mid\eta)\) and \(\mathrm{P}(d_i,S\mid\eta)\) numerically.

\begin{table}
\caption{Inference cases used to compare flux information, microlensing, and selection effects.}
\label{tab:inference_runs}
\centering
\renewcommand{\arraystretch}{1.15}
\begin{tabular}{c c >{\centering\arraybackslash}m{0.15\textwidth} c}
\hline
\hline
Case & Constraints & Flux model & Selection effect \\
\hline
A
& \(\theta_{1,2}^{\rm obs}, f_{1,2}^{\rm obs}\)
& Microlensing
& Included \\

B
& \(\theta_{1,2}^{\rm obs}, f_{1,2}^{\rm obs}\)
& Macromodel
& Included \\

C
& \(\theta_{1,2}^{\rm obs}\)
& Microlensing
& Included \\

D
& \(\theta_{1,2}^{\rm obs}\)
& Not applicable
& Not included \\

E
& \(\theta_{1,2}^{\rm obs}\)
& Macromodel
& Included \\

F
& \(\theta_{1,2}^{\rm obs}\)
& Macromodel with \(0.3\,{\rm dex}\) scatter
& Included \\
\hline
\end{tabular}
\tablefoot{
All cases also use \(m_{*,{\rm SPS}}^{\rm obs}\) and \(r_{\rm e}^{\rm obs}\) as constraints.
For Cases A and B, the flux model is used both to describe the observed image fluxes and to account for the selection effect.
For Cases C, E, and F, it is used only to account for the selection effect.
}
\end{table}

\subsection{Comparison inference cases}
We take the model described above as our standard inference model, which we refer to as Case A. To investigate how image fluxes affected by microlensing improve the constraints, how sensitive the constraints are to the choice of flux model, and how the results depend on the inclusion of the selection effect, we define five comparison inference cases in addition to Case A. All cases adopt the same foreground population model and source population model. They differ in three respects: whether the selection effect is included, whether the observed image fluxes are used in the likelihood, and which model is adopted for the physical image flux distribution, \(\mathrm{P}(\vec{f}\mid\psi)\). For all cases, the likelihood includes the image positions, the effective radius, and the SPS stellar mass. 

We consider three models for \(\mathrm{P}(\vec{f}\mid\psi)\). The microlensing model uses the microlensing magnification distribution. The macromodel predicts the image fluxes using only the absolute macromodel magnifications. The Gaussian model adds a scatter of \(0.3\) dex in \(\log f\) predicted by the macromodel.

The six cases are summarized in Table~\ref{tab:inference_runs}. In Case A, the microlensing flux model is used both to describe the observed image fluxes and to account for the selection effect. Case B also uses the observed image fluxes as constraints, but replaces the microlensing flux model with the macromodel. Cases C--F do not use the observed image fluxes as constraints. Case C uses the microlensing flux model to account for the selection effect, while Case D does not include the selection effect. Cases E and F account for the selection effect using the macromodel without and with a scatter of \(0.3\) dex, respectively.
 \section{Results}
\label{sec:results}

\begin{figure*}
    \centering
    \includegraphics[width=\textwidth]{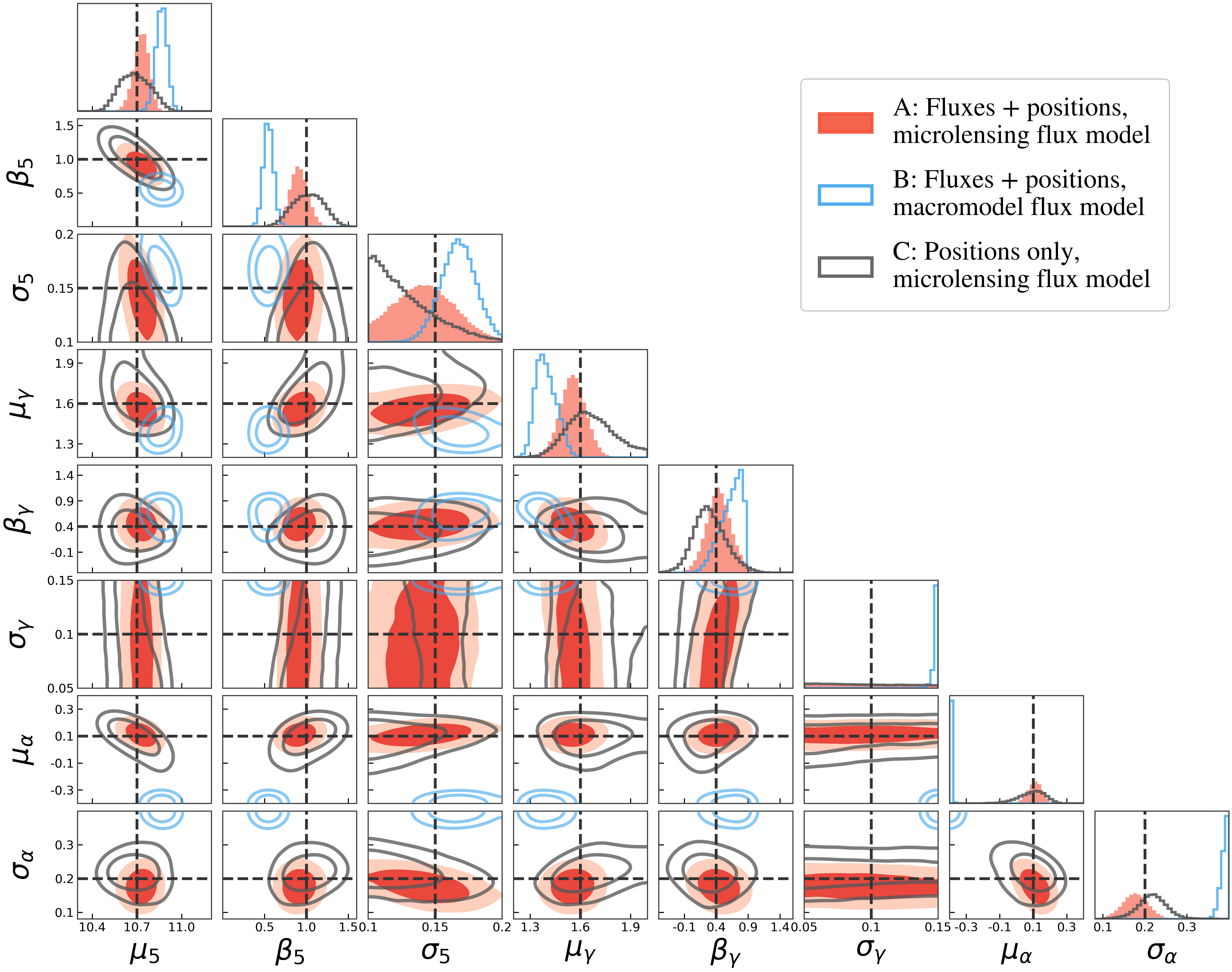}
    \caption{Posterior distributions of the foreground population hyperparameters for Cases A--C defined in Table~\ref{tab:inference_runs}. The constraints and flux models are indicated in the legend. In Cases A and B, the flux model is used both to describe the observed image fluxes and to account for the selection effect. In Case C, it is used only to account for the selection effect. Dashed lines mark the input hyperparameter values.}
    \label{fig:corner_compare_three_chains}
\end{figure*}

Figure~\ref{fig:corner_compare_three_chains} shows the posterior distributions of the population parameters \(\eta\) obtained by fitting the mock lens sample with the model described in Section~\ref{sec:inference}. The orange contours show the posterior for Case A, obtained using the observed image fluxes and image positions as constraints. The gray contours show the posterior for Case C, obtained using the image positions without the observed image fluxes. Both cases recover the input population parameters, but Case A gives tighter constraints. For example, the width of the 68\% credible interval on \(\mu_\alpha\) decreases from \(0.16\) dex in Case C to \(0.09\) dex in Case A. The improvement is clearest in the \(\mu_\alpha-\mu_5\) and \(\mu_\alpha-\mu_\gamma\) planes, where the microlensed fluxes help reduce the degeneracies among the population mean parameters \(\mu_\alpha\), \(\mu_5\), and \(\mu_\gamma\).

This is expected because image positions mainly constrain the total projected mass distribution of the lens. They do not uniquely determine how the mass is divided between the stellar component and the dark matter component. Microlensed image fluxes add information about the local stellar contribution at the image positions through their dependence on the local convergence, shear, and smooth matter fraction.

The blue contours show the posterior for Case B, obtained using the observed image fluxes and image positions with the macromodel flux model (i.e., without accounting for microlensing). The posterior shifts away from the input hyperparameter values. In particular, \(\mu_\alpha\) approaches its lower prior boundary, while \(\sigma_\gamma\) approaches its upper prior boundary. This bias occurs because the model absorbs the microlensing perturbations in the image fluxes into the foreground population parameters. Thus, image fluxes provide reliable population information only when microlensing is included in the inference model. The corresponding numerical constraints are listed in Table~\ref{tab:inference}.

\begin{table*}
\caption{Inference on the foreground galaxy population hyperparameters.}
\label{tab:inference}
\begin{tabularx}{\textwidth}{ccccccX}
\hline
\hline
Parameter & Truth & Prior & A & B\(^{\dagger}\) & C & Description \\
\hline
$\mu_5$ & $10.70$ & $U(10.20,11.20)$ & $10.73^{+0.05}_{-0.05}$ & $10.87^{+0.04}_{-0.04}$ & $10.68^{+0.11}_{-0.10}$ & Mean $m_5$ at $m_{*,\rm SPS}=10.9$ \\
$\beta_5$ & $1.00$ & $U(0.00,2.00)$ & $0.91^{+0.11}_{-0.10}$ & $0.55^{+0.06}_{-0.06}$ & $1.02^{+0.18}_{-0.19}$ & Scaling of mean $m_5$ with $m_{*,\rm SPS}$ \\
$\sigma_5$ & $0.15$ & $U(0.10,0.50)$ & $0.14^{+0.02}_{-0.02}$ & $0.17^{+0.01}_{-0.01}$ & $0.12^{+0.03}_{-0.02}$ & Scatter in $m_5$ around the mean \\
$\mu_\gamma$ & $1.60$ & $U(1.10,2.10)$ & $1.56^{+0.08}_{-0.08}$ & $1.39^{+0.07}_{-0.05}$ & $1.66^{+0.17}_{-0.13}$ & Mean $\gamma_{\rm DM}$ at $m_{*,\rm SPS}=10.9$ \\
$\beta_\gamma$ & $0.40$ & $U(-2.60,3.40)$ & $0.44^{+0.19}_{-0.19}$ & $0.66^{+0.14}_{-0.21}$ & $0.28^{+0.27}_{-0.23}$ & Scaling of $\gamma_{\rm DM}$ with $m_{*,\rm SPS}$ \\
$\sigma_\gamma$ & $0.10$ & $U(0.05,0.15)$ & $0.10^{+0.03}_{-0.03}$ & $0.15^{+0.00}_{-0.00}$ & $0.09^{+0.04}_{-0.03}$ & Scatter in $\gamma_{\rm DM}$ around the mean \\
$\mu_\alpha$ & $0.10$ & $U(-0.40,0.60)$ & $0.11^{+0.04}_{-0.04}$ & $-0.40^{+0.01}_{-0.00}$ & $0.10^{+0.07}_{-0.09}$ & Mean $\log\alpha_{\rm SPS}$ \\
$\sigma_\alpha$ & $0.20$ & $U(0.10,0.50)$ & $0.18^{+0.03}_{-0.03}$ & $0.40^{+0.02}_{-0.01}$ & $0.22^{+0.03}_{-0.03}$ & Scatter in $\log\alpha_{\rm SPS}$ around the mean \\
\hline
\end{tabularx}
\tablefoot{
Posterior entries give the median and the central 68\% credible interval.
The dagger marks a case in which at least one parameter remains close to a prior boundary after convergence in all chains.
}
\end{table*} 
\begin{figure*}
    \centering
    \includegraphics[width=\textwidth]{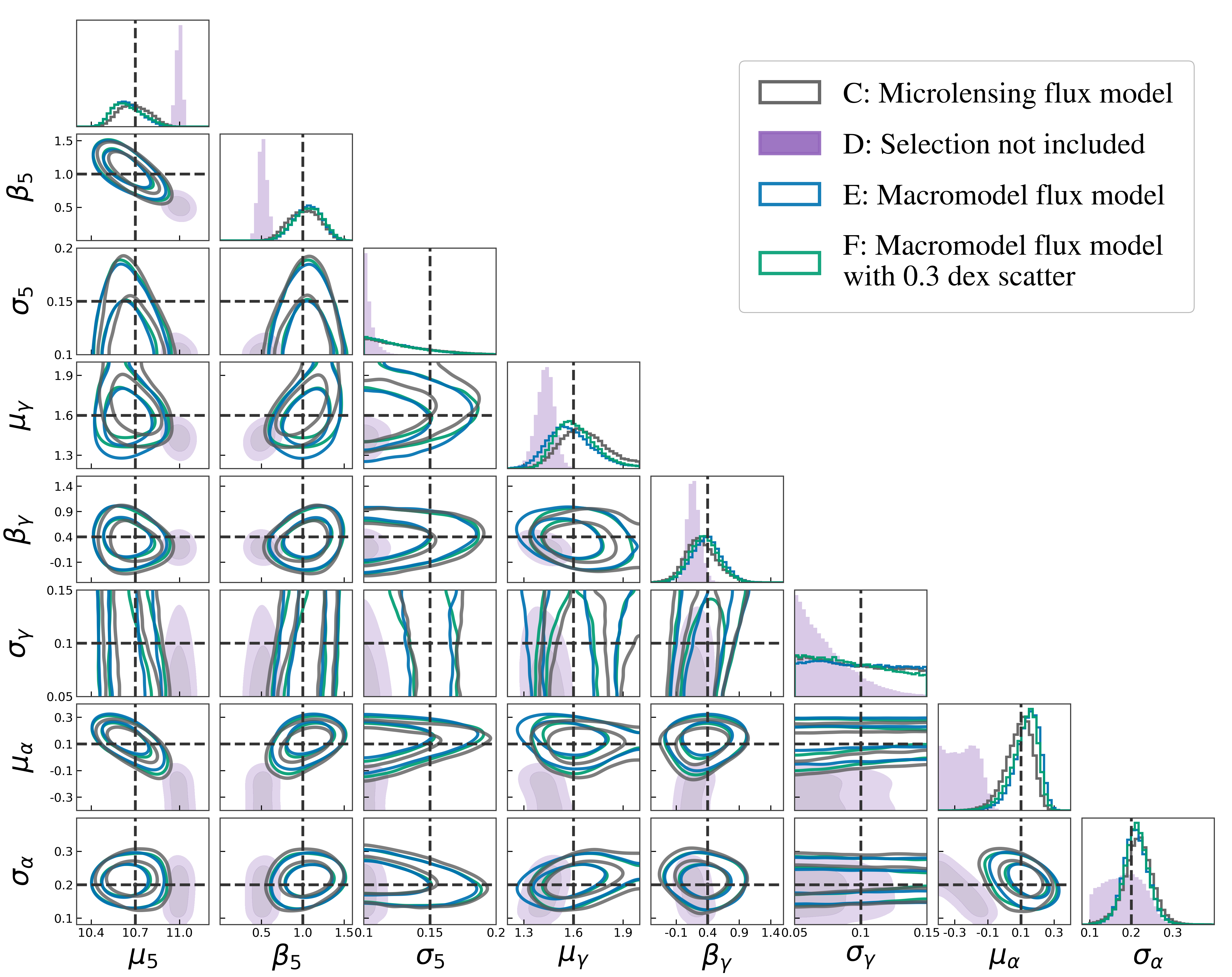}
    \caption{Posterior distributions of the foreground population hyperparameters for Cases C--F defined in Table~\ref{tab:inference_runs}. None of these cases uses the observed image fluxes as constraints. In Cases C, E, and F, the flux model indicated in the legend is used only to account for the selection effect, while Case D does not include the selection effect. Dashed lines mark the input hyperparameter values.}
    \label{fig:corner_compare_ip_selection}
\end{figure*}

Figure~\ref{fig:corner_compare_ip_selection} compares Cases C--F, none of which use the observed image fluxes as constraints. Case C accounts for the selection effect using the microlensing flux model, while Case D does not include the selection effect. Case E uses the macromodel to account for the selection effect, and Case F uses the macromodel with a Gaussian scatter of \(0.3\) dex.

Cases E and F give posterior constraints similar to Case C, and all three recover the input population parameters. In contrast, Case D gives biased constraints, with the largest shifts in \(\mu_5\), \(\beta_5\), and \(\mu_\alpha\). For the mock sample considered here, this shows that, when the observed fluxes are not modeled directly and the selection function depends on flux, the inference is sensitive to whether the selection effect is included, but not strongly sensitive to the flux perturbation model. The corresponding numerical constraints are listed in Table~\ref{tab:selection_inference}.

\begin{table*}
\caption{Inference on the foreground galaxy population hyperparameters for the selection comparison.}
\label{tab:selection_inference}
\begin{tabularx}{\textwidth}{cccccccX}
\hline
\hline
Parameter & Truth & Prior & C & D & E & F & Description \\
\hline
$\mu_5$ & $10.70$ & $U(10.20,11.20)$ & $10.68^{+0.11}_{-0.10}$ & $11.00^{+0.02}_{-0.02}$ & $10.64^{+0.10}_{-0.09}$ & $10.64^{+0.12}_{-0.09}$ & Mean $m_5$ at $m_{*,\rm SPS}=10.9$ \\
$\beta_5$ & $1.00$ & $U(0.00,2.00)$ & $1.02^{+0.18}_{-0.19}$ & $0.52^{+0.06}_{-0.05}$ & $1.07^{+0.16}_{-0.17}$ & $1.07^{+0.17}_{-0.18}$ & Scaling of mean $m_5$ with $m_{*,\rm SPS}$ \\
$\sigma_5$ & $0.15$ & $U(0.10,0.50)$ & $0.12^{+0.03}_{-0.02}$ & $0.10^{+0.01}_{-0.00}$ & $0.12^{+0.03}_{-0.02}$ & $0.12^{+0.03}_{-0.02}$ & Scatter in $m_5$ around the mean \\
$\mu_\gamma$ & $1.60$ & $U(1.10,2.10)$ & $1.66^{+0.17}_{-0.13}$ & $1.43^{+0.05}_{-0.06}$ & $1.57^{+0.16}_{-0.12}$ & $1.60^{+0.14}_{-0.11}$ & Mean $\gamma_{\rm DM}$ at $m_{*,\rm SPS}=10.9$ \\
$\beta_\gamma$ & $0.40$ & $U(-2.60,3.40)$ & $0.28^{+0.27}_{-0.23}$ & $0.17^{+0.11}_{-0.09}$ & $0.38^{+0.23}_{-0.24}$ & $0.34^{+0.23}_{-0.23}$ & Scaling of $\gamma_{\rm DM}$ with $m_{*,\rm SPS}$ \\
$\sigma_\gamma$ & $0.10$ & $U(0.05,0.15)$ & $0.09^{+0.04}_{-0.03}$ & $0.07^{+0.03}_{-0.01}$ & $0.10^{+0.04}_{-0.03}$ & $0.09^{+0.04}_{-0.03}$ & Scatter in $\gamma_{\rm DM}$ around the mean \\
$\mu_\alpha$ & $0.10$ & $U(-0.40,0.60)$ & $0.10^{+0.07}_{-0.09}$ & $-0.25^{+0.10}_{-0.10}$ & $0.14^{+0.06}_{-0.08}$ & $0.13^{+0.06}_{-0.09}$ & Mean $\log\alpha_{\rm SPS}$ \\
$\sigma_\alpha$ & $0.20$ & $U(0.10,0.50)$ & $0.22^{+0.03}_{-0.03}$ & $0.18^{+0.05}_{-0.05}$ & $0.21^{+0.03}_{-0.03}$ & $0.21^{+0.03}_{-0.03}$ & Scatter in $\log\alpha_{\rm SPS}$ around the mean \\
\hline
\end{tabularx}
\tablefoot{
Posterior entries give the median and the central 68\% credible interval.
}
\end{table*}
  \section{Discussion}
\label{sec:discussion}

Our analysis suggests that, by applying our methodology to a sample of order 200 lenses, we can constrain the mean $\log\alpha_{\rm SPS}$ with an uncertainty of approximately $0.05$ dex ($12\%$). This would constitute a major step forward compared to the current uncertainty, which is approximately a factor of two \mbox{\citep{smithEvidenceInitialMass2020a}}.
However, our results are based on a set of simplifying assumptions.
The most important one is the assumption of circular symmetry in the mass distribution of the lenses.
In reality, galaxies are not circularly symmetric, and so a more complex model must be adopted in order to fit real samples of lenses. 
The azimuthal structure of a lens is degenerate with its radial density profile \mbox{\citep{kochanekOverconstrainedModelsTime2021}}, and the positions of two lensed images alone cannot determine the ellipticity of a lens with high precision \mbox{\citep{bradyTDCOSMOXXVIUniform2026}}. This means that, with a more realistic version of our method, we expect the precision to decrease compared to the results reported in this paper.
Such a drop in precision could be mitigated in practice by the inclusion of quadruply-imaged lenses, and by including constraints from the full surface brightness distribution of the lensed galaxy that is hosting the quasar.
Nevertheless, the goal of this work is to demonstrate the accuracy in the recovery of the galaxy population parameters. Precision may be gained by increasing the size to which our method is applied.

Second, the modelling of image fluxes needs to be extended beyond the simplified microlensing model adopted here. We used the point source approximation, for which the microlensing magnification distribution depends only on the local lensing quantities. This approximation is appropriate for very compact emission regions, such as the region that is responsible for the X-ray emission, but many available quasar flux measurements are obtained in optical, infrared, or radio bands, where the source size can be significant. A more general treatment requires microlensing magnification distributions that depend on the source surface brightness profile and the microlens mass function. Therefore, applying this framework to future large surveys such as Euclid, CSST, and Roman, particularly for optical and near-infrared quasar observations, requires a prior model for the quasar surface brightness distribution and a realistic microlens mass function in lens galaxies. For applications to real data, the surrogate microlensing magnification model adopted in this work will need to be replaced. A computationally efficient tool for evaluating finite-source microlensing magnification distributions will also be required to enable their incorporation into hierarchical inference. Developing such a more general treatment is beyond the scope of this work and will be pursued in future studies.

In addition, microlensing may not be the only source of flux perturbations in real lens systems. Observed flux ratios can also be affected by dark matter subhalos, line-of-sight structures, dust extinction, intrinsic quasar variability combined with time delays, and uncertainties in the macroscopic lens model. In this work, we assume that microlensing dominates these flux perturbations. Whether these effects need to be considered and how they should be incorporated into the flux model will require further investigation for applications to real data.

Finally, applying this framework to real data requires realistic models for both the lens galaxy and source quasar populations. In this work, the lens galaxy population is described by simplified distributions used to generate the mock sample. For real applications, the distributions of lens galaxy properties, such as effective radius and stellar mass, can be obtained from observations of appropriate parent galaxy samples. In addition, lens redshifts need to be measured for individual systems, and the joint distribution of galaxy properties and redshift needs to be characterized. The source population requires a model for the intrinsic quasar luminosity function and redshift distribution. These models can be obtained from large quasar surveys. However, the quasar luminosity function needs to be well measured at the faint end relevant for the selected lensed population, since magnification bias depends on the intrinsic source flux distribution.

 \section{Conclusions}
\label{sec:conclusion}

We have developed a hierarchical framework for using microlensed quasars to infer the population properties of lens galaxies.
The method combines image position constraints with a flux likelihood in which the image fluxes are modeled using microlensing magnification distributions.
We tested the method on a simplified mock sample of doubly imaged quasars. Our results show that microlensed image fluxes can provide additional
constraints on the population properties of lens galaxies when microlensing
is properly accounted for. In particular, the flux information helps us to break
the degeneracy between the stellar and dark matter components that remains
when using image positions alone. This suggests that microlensed fluxes
from large samples of lenses with a reproducible selection function can provide a practical way of distinguishing between different IMFs at the population level.

We also found that an incorrect image flux model can lead to severely biased constraints, as demonstrated by the case where microlensing effects are ignored. In addition, neglecting the selection effect can introduce significant biases in the inferred population parameters. However, when image fluxes are not used as direct constraints, the choice of flux model used only to describe the selection function does not significantly affect the inference results in the simplified mock considered here.

Applying this method to real data will require dropping the assumption of circular symmetry in the lens model, and to improve the fidelity of the microlensing magnification model.
Nevertheless, this work represents a first necessary step towards the full exploitation of a statistical sample of strongly lensed quasars.
 
\bibliographystyle{aa}

\bibliography{StatVNew}

\end{document}